\documentclass[preprint,showpacs,showkeys,preprintnumbers,amsmath,amssymb,superscriptaddress]{revtex4}
\usepackage{graphicx}
\usepackage{dcolumn}
\usepackage{bm}
\usepackage[dvips]{color}
\usepackage{epsfig}

\begin{document}

\title{Emergent GUP from Modified Hawking Radiation in Einstein-NED Theory}

\author{S. Hamid Mehdipour}

\email{mehdipour@liau.ac.ir}

\affiliation{Department of Physics, College of Basic Sciences,
Lahijan Branch, Islamic Azad University, P. O. Box 1616, Lahijan,
Iran}

\date{\today}

\begin{abstract}
We present a general procedure for constructing exact black hole
(BH) solutions with the magnetic charge in the context of nonlinear
electrodynamics (NED) theory as well as in the coherent state
approach to noncommutative geometry (NCG). In this framework, the
Lagrangian density for the noncommutative Hayward BH is obtained and
the weak energy condition (WEC) is satisfied. The noncommutative
Hayward solution depends on two kind of charges which for the
vanishing of them yields the Schwarzschild solution. Moreover, in
order to find a link between the BH evaporation and uncertainty
relations, we may calculate the Hawking temperature and find the
effect of the Lagrangian density of BHs on the Hawking radiation.
Therefore, a generalized uncertainty principle (GUP) emerges from
the modified Hawking temperature in Einstein-NED theory. The origin
of this GUP is the combined influence of a nonlinear magnetic source
and an intrinsic property of the manifold associated with a
fictitious charge. Finally, we find that there is an upper bound on
the Lagrangian uncertainty of the BHs which are sourced by the NED
field and/or the fictitious charge.
\end{abstract}


\maketitle

\section{\label{sec:1}Introduction}
It has been shown that the physical source of a regular black hole
(RBH) can be interpreted as the gravitational field of a nonlinear
electrodynamics (NED) \cite{pell,bronn,bronn2,ayo1,ayo2,ayo3}. NED
models emerge from the low-energy effective limit in particular
approaches to string/M-theories \cite{frad,tsey,seib}. There are two
important purposes in a NED theory. One is to take into account
electromagnetic field and particles in the frame of a physical
source and the other is to ignore allowing physical quantities turn
into infinite. A same process is attained by a physically reliable
NED model coupled to gravity so that regular spherically symmetric
electrically charged solutions satisfy the weak energy condition
(WEC) and contain an inescapable deSitter centre of the nonlinear
electrodynamic source \cite{ans}. The metric and curvature
invariants of charged RBHs, incompatible with Reissner-Nordstr\"{o}m
(R-N) BHs, are regular everywhere. The Bardeen BH is the first RBH
model in general relativity (GR) which has been proposed by the
pioneering work of Bardeen in 1968 \cite{bar}. This BH has an event
horizon and a deSitter-like feature inside its horizon which does
not violate the WEC. A bit later, Pelliger and Torrence \cite{pell}
obtained a general static, spherically symmetric solution with an
electric charge in GR coupled to the NED with a gauge-invariant
Lagrangian. Later, in 1976, Bronnikov and Shikin \cite{bronn} proved
in a general form that if this NED has a Maxwell weak field limit,
then the above solution cannot have a regular centre (see also
\cite{bronn2}). In 2001, Bronnikov \cite{bron} extended the solution
to include radial magnetic field and showed that only purely
magnetic configurations can have a regular centre, while the regular
metric "profile" in this case can be arbitrary. The Hayward solution
is another well-known kind of regular spacetime wherein its static
region is Bardeen-like while the dynamic regions are Vaidya-like
\cite{hay}. Ayon-Beato and Garcia reinterpreted the Bardeen BH as a
magnetic solution to Einstein equations with the NED \cite{ayo4}.
The other solutions of Einstein-NED theory have also been reported
in \cite{fan1,fan2,tosh}. There have been a great number of studies
concerning the combined Einstein and NED equations in the literature
\cite{ayo5,bron2,bere,azre,bala,radi,juni,dymn,chav,bronn4,bronn5,NED-recent}.

The issue of central singularity in BHs is widely believed to be an
unavoidable prediction of GR \cite{ell}. However, various
phenomenological approaches have been considered in the literature
for resolving the problem via a regular centre \cite{ans}. In a
noncommutative geometry (NCG) inspired model (for a review see
\cite{nic1}), a point particle in a noncommutative spacetime is no
longer characterized by a Dirac-delta function distribution, but
will be described as a smeared particle by a Gaussian distribution
of minimal width $\sqrt{\theta}$, beyond which coordinate resolution
is obscure. Therefore, the Einstein tensor in gravity field
equations stays unchanged however the energy-momentum tensor takes a
new form. As a striking result of this noncommutativity model, the
curvature singularity at the origin of BHs is removed. Instead of
the curvature singularity, a regular deSitter vacuum state will
appear concerning the effect of the strong quantum fluctuations at
short distances. In fact, a noncommutative BH is a combination of
the deSitter core around the origin with the ordinary metric of the
BH far away from the origin. So, the existence of a deSitter core in
the centre of a BH prevents its collapse into a singular state, and
an ordinary metric at large distances is recovered where the
behavior of the minimal length is not influential.

On the other hand, in recent years, due to various conceptual and
technical reasons, it has been suggested that there may exist
corrections to the Heisenberg uncertainty principle (HUP) which
could be important at extreme scales, i.e. in the ultraviolet (UV)
and in the infrared (IR) scales \cite{kem}. In the UV scale, the
incorporation of gravity in quantum field theory yields effectively
a cutoff in the high energy regime, i.e. a minimal length scale of
the order of the Planck length, $l_{P}\approx 10^{-35}m$. Studies on
string collisions at Planckian energies and through a
renormalization group type of analysis lead to the modification of
the HUP, i.e. the generalized uncertainty principle (GUP)
\cite{amat0} (for a review, see \cite{taw}). The thermodynamics of
Schwarzschild and R-N BHs using the GUP was studied in
\cite{Gangopadhyay}. The exact expressions for the mass-temperature
relation, heat capacity and entropy were obtained. They revealed
that the expression for the entropy is the well known area theorem
in terms of the horizon area up to leading order corrections from
the GUP. They also found that the the entropy has a singularity at
zero corrected horizon area that refers to a critical mass which is
less than the remnant mass at which the BH evaporation process
halts. For a review of the various roads to the remnant scenario and
their challenges for ameliorating the BH information loss problem
\cite{Page:1993}, see \cite{Chen:2015}.

The whole problems concerning the BH evaporation \cite{Hawking:1975}
are mostly owing to a failure of the semiclassical explanation under
the condition that the BH mass should be extremely larger than the
Planck mass. When the decay continues and the BH mass reduces, this
condition ultimately can no longer be true and a theory of quantum
gravity (QG) must be applied (for a review, see \cite{stro}).
Further, Parikh and Wilczek \cite{Parikh:2000} found that Hawking
radiation can be extracted from the null geodesic method by
considering the backreaction effects. They calculated a leading
correction to the probability of massless particles tunneling across
the horizon. The tunneling process illustrates an extended emitted
spectrum that is not definitely thermal but is consistent with an
underlying unitary quantum theory. In fact, the emission rate
deviates from the pure thermal spectrum and results in unitarity; so
that the conservation of information is supported
\cite{Parikh:2004}. There is another well-known tunneling approach
for exploring the Hawking radiation as a quantum tunneling effect
that is the Hamilton-Jacobi strategy \cite{Srinivasan}. Many authors
considered alternative derivations to compute the Hawking radiation
and also investigated the tunneling of various particles to obtain
thermodynamical quantities such as the temperature and entropy for
different kind of BHs \cite{tunneling}.

Applying the GUP modified entropy to the entropic gravity scenario
proposed by Verlinde \cite{Verlinde}, suggesting that the gravity is
originally an entropic force, exhibits that the resulting gravity
force law includes sub-leading order correction terms that is
dependent on the Planck constant \cite{Chen:2011}. Such modification
in the area law of the BH entropy could lead to a modified Newton's
law of gravitation and could have observable consequences at length
scales much larger than the Planck scale \cite{Ali:2013}.

Recently, Feng {\it et al.} \cite{Feng:2017} have demonstrated that
the dimensionless parameter in the GUP can be constrained by the
gravitational wave event GW150914 discovered by the LIGO Scientific
and Virgo Collaborations \cite{abb1}. They have derived the
energy-momentum dispersion relation and the difference between the
propagation speed of gravitons and the speed of light using the HUP
and the data of gravitational wave event. Moreover, they have
obtained the upper limits of the GUP parameters based on the
modified speed of gravitons. As another example of possible
experimental signatures of modified uncertainty principle, Mureika
\cite{Mureika:2019} proposed an extended uncertainty principle that
allows for large scale modifications to gravity. This would have an
impact on the attributes of most known supermassive BHs, and
therefore could be observed by the Event Horizon Telescope
\cite{EHT} or other future collaborations.

Here it is worth noting that there are also some comments on
opposite ideas on GUP. For example, GUP measurements could have an
effect on the violation of fundamental symmetries, like CPT and
Lorentz invariance in the framework of the standard model extension
\cite{Lambiase}. The basic reason for this problem, as mentioned
above, is that QG corrections such as those of the GUP serve no
purpose in the context of Hawking radiation. It was already evident
in the earliest literature \cite{Hawking} that the process of
Hawking emission is only valid for semiclassically large BHs. In
this case, such corrections would be undetectable leading-order
quantum effects, is debatable to begin with. The same requirement of
large BHs dooms the idea that anything can be learned about the end
point of BH evaporation in this program. The legitimacy of BH
thermodynamics has broken down long before the alleged remnants have
been identified. In fact, any method that does predict the existence
of remnants should be viewed upon with a large dose of skepticism.

Nevertheless, the application of the GUP to BH thermodynamics has
attracted considerable attention and leads to significant
modifications to the emission process, particularly at the final
stages of the evaporation \cite{blt}. It is worth mentioning, the
thermodynamics of BHs is thought to be the connection between BH
physics and quantum theory. For instance, in 2001, Adler {\it et
al.} \cite{adl} have argued that in opposition to standard
viewpoint, the GUP may prevent small BHs from total disappearing in
exactly the same way that the HUP obstructs the hydrogen atom from
total collapse. They applied the GUP approach in an alternative
heuristic derivation of the Hawking radiation to find a corrected BH
temperature. In this heuristic method, the Hawking temperature can
be achieved by the utilization of the HUP and general properties of
BHs \cite{oha}. Many authors considered various problems in this
framework, e.g., see \cite{gup}.

Several studies in string theory and NCG give a particular type of
correction to the HUP and thus propose the appearance of a finite
limit to the possible resolution of distances at extreme scales
\cite{amat}. Indeed, because of the appearance of extreme energies
at short distances of a noncommutative manifold, the influences of
manifold quantum fluctuations turn into noticeable and prevent any
measurements to determine a particle position with a preciseness
greater than an intrinsic length scale which can be understood as an
example of locality violation at the Planck length \cite{sma}. It is
interesting to note that the methodology of the NCG effectively
represents itself as an equivalent description of the nonlocal field
theory discussed in Ref.~\cite{eug}. It can be inferred from BH
physics that any theory of QG should have an intrinsic minimum
length of the order of Planck length \cite{mag}. There are some
other methods to investigate the QG effects of BHs that have been
studied in the literature \cite{Ali}.

On the basis of the noncommutativity, the extracting of metrics for
noncommutative BHs is identified with the possible running of the
minimal observable length in GR. Recently, we have analysed the
final stages of the BH evaporation for the noncommutative Bardeen
and Hayward solutions \cite{meh1,meh2}. The results showed that the
behavior of Hawking radiation changes considerably at the small
radii regime such that the BH does not evaporate completely, but a
stable BH remnant will be remained at the final phase of the
evaporation. In this paper, we are going to find the Lagrangian
density of a term depending nonlinearly on the electromagnetic field
tensor for a family of spherically symmetric, static, charged RBH
metrics. We obtain the Hawking temperature of the noncommutative
Hayward BH and use its corrected temperature to derive a modified
uncertainty relation including the Lagrangian density in exactly the
same way that the GUP creates a modified Hawking temperature.

This paper is organized as follows. In Sec.~\ref{sec:2}, we describe
the nonlinear electromagnetic field equations in regular spacetimes.
The noncommutative effects on the spacetime of Hayward are
investigated in Sec.~\ref{sec:3}. The resulting metric in three
possible causal structures is analyzed and its WEC is examined. In
Sec.~\ref{sec:4}, we determine the Hawking temperature of the
noncommutative Hayward BH and then, in order to find the effect of
the Lagrangian density on the Hawking radiation, we speak of a
modified uncertainty relation for radiated photons that is derived
from the HUP. Thus, a rough estimate between the Lagrangian density
of RBHs and uncertainty relations is revealed. Finally, our results
are briefly presented in Sec.~\ref{sec:5}. Throughout the paper,
Greek indices run from 0 to 3 and we use natural units with the
following definitions: $\hbar= c = G= k_B = 1$.

\section{\label{sec:2}NED field coupled with GR}
Many configurations of NED coupled to GR now exist in the literature
(for a brief review see \cite{bronn3}). In this section, we
investigate the most general form of a spherically symmetric,
static, charged RBH metrics in GR coupled to a NED. We start with
the action describing the dynamics of a self-gravitating NED field
in GR
\begin{equation}
\label{mat:1}S=\frac{1}{16\pi}\int\sqrt{-\textsf{g}}d^4x\left(R-L(F)\right),
\end{equation}
where $R$ is the curvature scalar with respect to the line element
$\textsf{g}_{\alpha\beta}$, and the Lagrangian density $L(F)$ is an
arbitrary nonlinear function of the Lorentz invariant, $F=
F_{\alpha\beta}F^{\alpha\beta}$, where
$F_{\alpha\beta}=\partial_\alpha A_\beta-\partial_\beta A_\alpha$ is
the electromagnetic field ($A_\alpha$ is the 4-vector potential). It
should be noted that the Lagrangian density approaches to Maxwell
asymptotic at weak electromagnetic fields, i.e. $L(F)\rightarrow F$,
and $L_F\equiv\frac{dL}{dF}\rightarrow1$ as $F\rightarrow 0$.

The tensor field $F_{\alpha\beta}$ satisfies equations
\begin{equation}
\label{mat:2}\nabla_\alpha\left(L_FF^{\alpha\beta}\right)=0,
\end{equation}
\begin{equation}
\label{mat:3}\nabla_\alpha \,^\ast\, F^{\alpha\beta}=0,
\end{equation}
where the asterisk refers to the Hodge dual. The stress-energy
tensor $T_{\alpha\beta}$ can be found by differentiating the action
$S$ with respect to the metric tensor $\textsf{g}_{\alpha\beta}$ as
follows
\begin{equation}
\label{mat:4}T_{\alpha\beta}=\frac{1}{4\pi}\left(L_FF_{\gamma\alpha}F^\gamma_\beta-\frac{1}{4}\textsf{g}_{\alpha\beta}L\right).
\end{equation}
The Einstein equations of motion are given by
\begin{equation}
\label{mat:5}R_{\alpha\beta}-\frac{1}{2}\textsf{g}_{\alpha\beta}R=2\left(L_FF_{\gamma\alpha}F^\gamma_\beta-\frac{1}{4}\textsf{g}_{\alpha\beta}L\right).
\end{equation}
Here, we consider a general ansatz of the BH solution having the
static spherical symmetric configuration with the nonlinear magnetic
charge
\begin{equation}\label{mat:6}  \Bigg\{
\begin{array}{ll}
ds^2=N(r)dt^2- N^{-1}(r)dr^2-r^2
(d\vartheta^2+\sin^2\vartheta d\phi^2),\\
A=e_m\cos\vartheta d\phi,\\
\end{array}
\end{equation}
where $e_m$ is the total magnetic charge carried by the BH related
to the value of Hayward's free parameter $g$ (given in length unit).
The value of $g$ is limited to the positive number, because the BH
solution only exists at the positive value. The metric function
$N(r)$ is defined by the relation
\begin{equation}
\label{mat:7}N(r)=1-\frac{2m(r)}{r}.
\end{equation}
Einstein's field equations may be simplified to these independent
forms
\begin{equation}
\label{mat:8}\frac{d^2N}{dr^2}-\frac{2}{r^2}(N-1)-\frac{4e_m^4}{r^4}L_F=0,
\end{equation}
\begin{equation}
\label{mat:9}\frac{1}{r}\frac{dN}{dr}+\frac{N-1}{r^2}+\frac{L}{2}=0.
\end{equation}
It can be seen that Eq.~(\ref{mat:8}) is satisfied for any given
metric function (\ref{mat:7}), arising from the most general form of
a static spherically symmetric solution with a magnetic charge.
Hence, using Eqs.~(\ref{mat:7}) and (\ref{mat:9}), the mass
distribution function can be written in the form
\begin{equation}
\label{mat:10}m(r)=\frac{1}{4}\int^rL\left(F(r')\right)r'^2dr'+C,
\end{equation}
where $C$ is an integration constant. In order to obtain $m(r)$, it
should be emphasized that as the equations of motion are too
complicated to solve analytically for dyonic charges, the situation
turns into much simpler for either magnetic or electric charges. In
this paper, we are going to construct exact BH solutions with the
magnetic charge because in the NED, the reason for regularity comes
from the magnetic charge. In a purely magnetic configuration,
$F_{\alpha\beta}$ is zero except for $F_{\vartheta\phi}$, that is
\begin{equation}
\label{mat:11}F_{\vartheta\phi}=e_m\sin\vartheta.
\end{equation}
Thus, using Eq.~(\ref{mat:2}), we find the square of the field
strength as
\begin{equation}
\label{mat:12}F=\frac{2e_m^2}{r^4}.
\end{equation}
The lagrangian density as a function of $r$, i.e.
$L=\frac{4}{r^2}\frac{dm}{dr}$, can be derived freely by selecting a
physically appropriate mass function to create static solutions
including magnetic charges. The ADM mass for a charged RBH has two
terms, one from the Schwarzschild mass and the other from the
nonlinear interactions between the graviton and the nonlinear
photon. The latter is not possible for a linear Maxwell field. It is
easily seen that the condition for the ADM mass at infinity leads to
$m(r\rightarrow\infty) = M = C$ (or $L=0$), and the Schwarzschild BH
is recovered as expected, which implies the solution of vacuum
Einstein equations. In addition, for $m(r)=M-\frac{e_m^2}{2r}$ (or
$L = F$), which implies the solution of Einstein-Maxwell theories,
we recover a magnetically charged R-N BH.

We remind that the causal structures of RBHs are similar to the R-N
BH, but with a deSitter centre instead of the singularity at $r =
0$. The breakdown of the classical picture takes place only in a
limited region around the origin, the region where gravity switches
to a repulsive quantum interaction. This is notably different from
the standard cases such as the Schwarzschild and R-N BHs which in
general have a singularity at the origin.

In NED, no singularity exists in a RBH. This is because the source
of RBHs is the nonlinear electrodynamic charge which differs from
the Schwarzschild case that has a mass shrinking to the singularity
as a result of losing mass by the emission of Hawking radiation. The
early stage of the Hawking evaporation process for the RBH is
similar to the Schwarzschild case, while for the late stage it is
wholly disparate. The Schwarzschild BH loses energy by the thermal
emission, with a gradual reduction in its mass and a growth on its
temperature. The occurrence of a divergent behavior of the Hawking
temperature is indeed due to the curvature singularity at the origin
of the Schwarzschild manifold. However, one can expect that the
evaporating BH will experience a Plank phase of the spacetime
manifold in the vicinity of the origin; so that the evaporation is
dramatically disturbed by strong quantum gravitational fluctuations
in which a QG theory must be employed.

\section{\label{sec:3}Noncommutative Hayward BH}
In this section, we are going to analyse the effect of an extended
structure associated with the microscopic discretion of spacetime on
the Hayward BH. We plan to consider the noncommutativity to have an
intrinsic minimum length scale equal to $\sqrt{\theta}$. Thus, a
point like structure will no longer be characterized by a
distribution which behaves like Dirac-delta function, but it will be
smeared by a distribution of minimal width $\sqrt{\theta}$. Here,
$\theta$ can be considered the smallest fundamental cell of an
observable area in the deformed theory. Even though the Einstein
tensor will not directly get deformed, the deformation of the
energy-momentum tensor by an extended structure will induces a
deformation of the original Einstein equation. The ordinary
classical metric will be recovered at large distances, but new
physics will be obtained at short distances, where the effect of an
extended structure cannot be neglected.

In the coordinate coherent states approach proposed by Smailagic and
Spallucci \cite{sma}, a point-like mass $M$ instead of being quite
localized at a point, is characterized by a smeared structure
throughout a region of linear size $\sqrt{\theta}$. The technique we
choose here is to seek to a static, spherically symmetric, minimal
width, Gaussian distribution of mass whose the noncommutative size
is determined by the parameter $\sqrt{\theta}$. Therefore, we should
model the mass distributions by a smeared delta function
\begin{equation}
\label{mat:13}\rho_{\theta}(r)={M\over
{(4\pi\theta)^{\frac{3}{2}}}}e^{-\frac{r^2}{4\theta}}.
\end{equation}
The matter density in Eq.~(\ref{mat:13}) displays a physical source
which is as close as it is possible to a point-like object. As has
been shown in Ref.~\cite{meh2}, the noncommutative Hayward mass
function $m(r)$ in terms of $g$ and $\theta$ in the metric function
(\ref{mat:7}) can be written as
\begin{equation}
\label{mat:14}m(r) =\frac{m_gm_\theta}{M},
\end{equation}
where the Hayward mass function $m_g$ in terms of $g$ and the
noncommutative mass function $m_\theta$ in terms of $\theta$ are
\begin{equation}\label{mat:15}  \Bigg\{
\begin{array}{ll}
m_g =M\left(\frac{r^3}{r^3+g^3}\right),\\
m_\theta =M\left[{\cal{E}}\left(\frac{r}{2\sqrt{\theta}}\right)
-\frac{r}{\sqrt{\pi\theta}} \textmd{e}^{-\frac{r^2}{4\theta}}\right],\\
\end{array}
\end{equation}
where the Gaussian error function is defined by ${\cal{E}}(x)\equiv
\frac{2}{\sqrt{\pi}}\int_{0}^{x}e^{-t^2}dt$. In the commutative
limit and $g\neq0$, the function $m(r)$ tends to the Hayward mass
function, i.e. $m\rightarrow m_g$, while for $\theta\neq0$ and
$g=0$, we get $m\rightarrow m_\theta$. This means that for
$\theta\rightarrow0$ and $g\rightarrow0$, the mass term defined in
Eq.~(\ref{mat:14}) is the same as the ADM mass at asymptotic
infinity, i.e. $M$, and this reduces to the Schwarzschild case.

\begin{figure}[htp]
\begin{center}$
\begin{array}{cccc}
\includegraphics[width=75 mm]{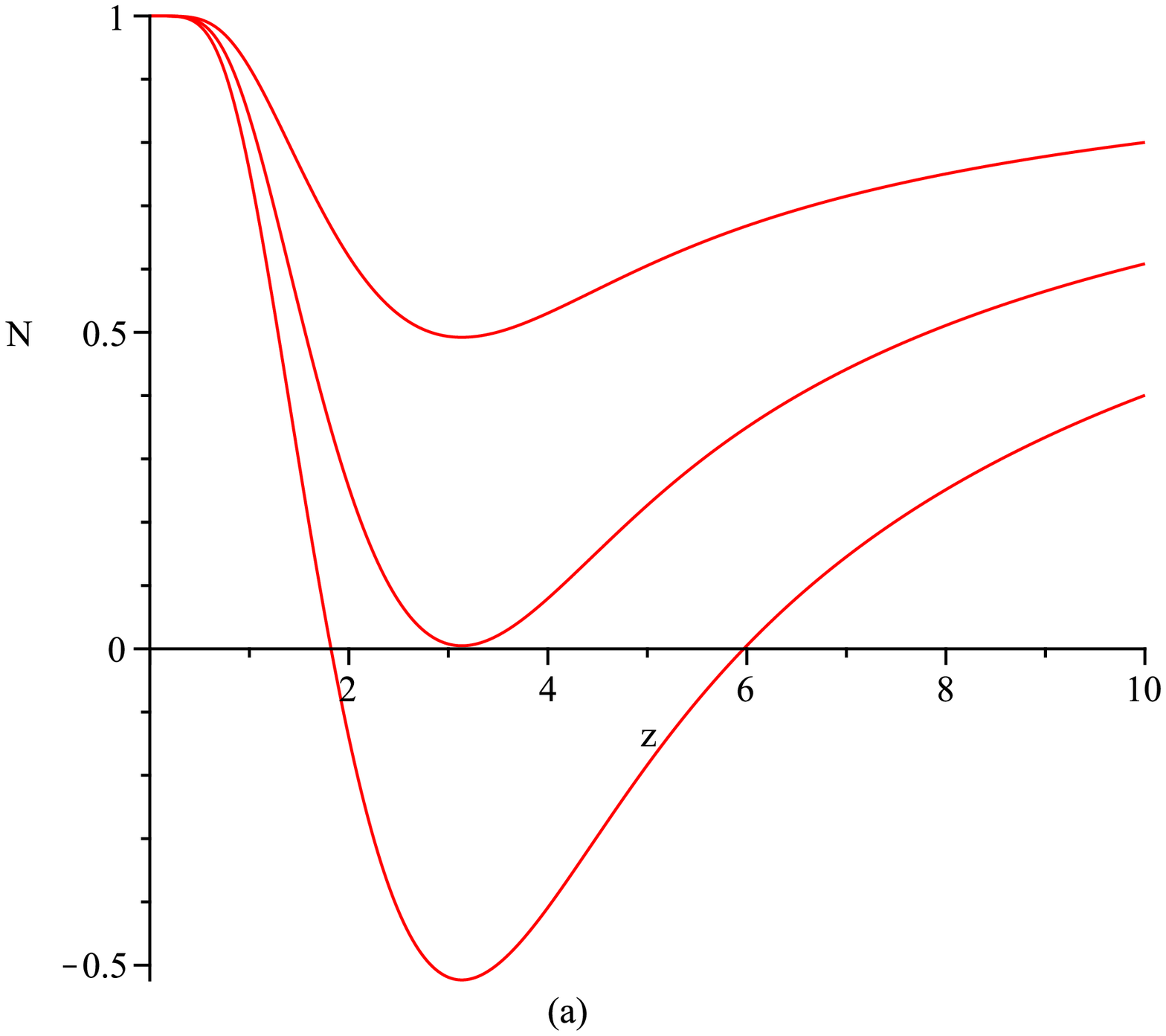}&\includegraphics[width=75 mm]{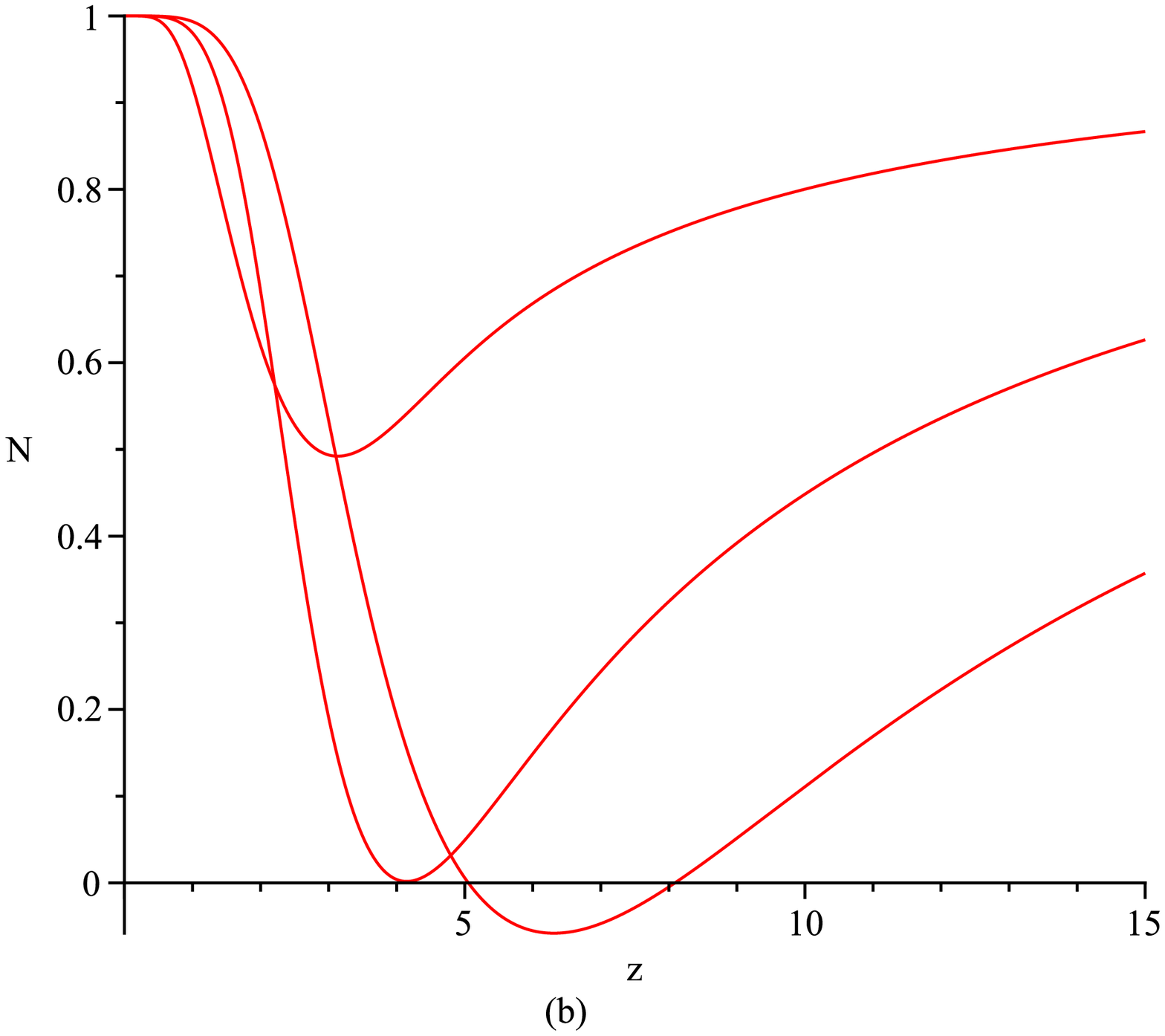}\\
\end{array}$
\end{center}
\caption{\scriptsize {$N$ in terms of $z$ for various values of $a$
and $s$. Figures ((a) and (b)) display three possible causal
structures: two distinct horizons, one degenerate horizon and no
horizon. In figure (a) we have set $a=1.00$. On the right-hand side
of figure (a), curves are marked from top to bottom by $s=0.50$ (no
horizon), $s=0.25$ (one degenerate horizon) and $s=0.17$ (two
distinct horizons) respectively. In figure (b) we have set $s=0.50$.
On the right-hand side of figure (b), curves are marked from top to
bottom by $a=1.00$ (no horizon), $a=22.42$ (one degenerate horizon)
and $a=125.00$ (two distinct horizons) respectively.  }}
 \label{fig:1}
\end{figure}

An interesting feature of the solution (\ref{mat:7}) is the horizon
equation $N(r_h) = 0$. One can draw plots and study the occurrence
of horizons. To this purpose, it is convenient to introduce the
dimensionless quantities $z=\frac{r}{\sqrt{\theta}}$,
$s=\frac{g}{2M}$ and $a=\frac{g^3}{\theta^{\frac{3}{2}}}$. Therefore
the metric function in terms of $z$ turns out to be
\begin{equation}
\label{mat:16}N(z)=1-\frac{a^{\frac{1}{3}}z^2\left[{\cal{E}}\left(\frac{z}{2}\right)
-\frac{z}{\sqrt{\pi}} \textmd{e}^{-\frac{z^2}{4}}\right]}{s(z^3+a)}.
\end{equation}
The numerical results of $N(z)$ versus $z$ are presented in
Fig.~\ref{fig:1}. Depending on the values of $a$ and $s$, the metric
function (\ref{mat:16}) displays different causal structures: the
entity of two horizons, one horizon (an extremal BH) and no horizon,
as can be seen from Fig.~(\ref{fig:1}).

It is important to note that for $z < z_0=\frac{r_0}{\sqrt{\theta}}$
one cannot speak of an event horizon and no temperature can be
defined. The physical description of $r_0$ is the smallest radius
which cannot be probed by a test particle that is located within
some distance from the source, so for $r < r_0$ we encounter an
unusual dynamical feature leading to, e.g., a negative temperature
\cite{meh3}. In fact, when the BH reaches the extremal configuration
with a minimal radius, the temperature is zero and the Hawking
emission abruptly stops. This means that, instead of the ordinary
divergent treatment for the ultimate phase of the Hawking
evaporation at small radii, there exists a value at which the
temperature vanishes. For that reason, the final phase of the BH
evaporation can be considered an extremal BH relic.

According to Fig.~(\ref{fig:1}), there are two distinct event
horizons for $z>z_0$. The existence of a minimal nonzero radius,
corresponding to the case of an extremal BH configuration ($z=z_0$),
is clear and hence for $z <z_0$ there is no event horizon, so that
there cannot be a BH. As figure (a) shows, the distance between the
horizons increases with decreasing the parameter $s$, while in
figure (b), as $a$ increases the distance between two horizons is
enlarged. Also, figure (b) implicitly indicates that the minimal
nonzero horizon radius increases with raising the parameter $a$. We
see that the causal structure is more sensitive to the values of
$s$, but is roughly unaffected by the values of $a$. This can be
rationally explained by the fact that the dimensionless quantity $s$
is dependent on the ADM mass $M$, which plays a key role in the BH
solutions at a broad range of scales, from the short distances to
the large distances, while the parameter $a$ is dependent on the
parameters which are more important at the short scale gravity.

The corresponding Lagrangian density for the noncommutative Hayward
BH in terms of $g$ and $\theta$ takes the form
\begin{equation}
\label{mat:17}L(r)=2m_g
\left(\frac{\textmd{e}^{-\frac{r^2}{4\theta}}}{\sqrt{\pi\theta^3}}+\frac{6g^3}{r^6}\frac{m}{M}\right).
\end{equation}
In the limit $\theta\rightarrow0$, one obtains the Lagrangian
density for the Hayward BH
\begin{equation}
\label{mat:18}L(r)=\frac{12g^3}{r^6}\frac{m_g^2}{M},
\end{equation}
and when $g\rightarrow0$, we have the Lagrangian density for the
noncommutative Schwarzschild BH
\begin{equation}
\label{mat:19}L(r)=\frac{2M\textmd{e}^{-\frac{r^2}{4\theta}}}{\sqrt{\pi\theta^3}}=16\pi\rho_{\theta}(r).
\end{equation}
The role of the parameter $\theta$ is pivotal because it is
responsible for the modified causal structure of the solution
(compared to the Schwarzschild case) which is similar to the R-N
case. For this reason we may call $\theta$ to \cite{nico}.

There is an important issue here which is worthwhile for discussion.
Assuming that the physical treatment of a RBH is qualitatively
comparable with or without noncommutativity, a key question has to
be asked: is there any motivation to propose the noncommutative's
parameter (the fictitious charge) accompanied by the RBH's parameter
(the magnetic charge)? In order to provide an accurate answer to
this question, it is necessary to point out that there are
fundamental distinctions between two situations. First, the
noncommutativity is not dependent on the curvature, but is an
inherent property of the manifold itself even in the absence of
gravity which is denoted by the parameter $\theta$ and can remove
some kind of divergences which appear in GR. Thus, if any effect is
made by the noncommutativity it must be revealed also in weak
fields. Second, the notion of weak or strong field is reasonable
only if we compare the field strength with a suitable scale. In GR,
there is a natural and unique scale, that is the Planck scale.
Hence, the gravitational field strength can still be regarded as
weak even near a BH, regarding the Planck scale. This justifies the
utilization of linearized field equations as a temporary laboratory
to determine the effect of noncommutativity until the horizon radius
is larger than the Planck length \cite{nico2}.

On the other hand, the Hayward BH is a regular solution of a
modified Einstein equation, and it is also found in a NED field
coupled with GR. In the near horizon of a BH, quantum effects are
significant due to the strong gravity, thus the spacetime geometry
may be modified from quantum effects close to the horizon and the
inherent singularity inside the BH may be removed. With this in
mind, we may imagine that the BH metric is modified close to the
horizon region owing to quantum effects. The range of deviations
from the standard solution of Einstein equations is determined by
free parameters of RBHs. Therefore, free parameters can explain how
much quantum effects nearby the horizon influence the deviation from
the standard energy level and the radiation. Along this line of
reasoning we take the charges as two different issues.

Here, one can expect that a physically reliable NED theory cannot
contravene the WEC. Despite the violation of the strong energy
condition to have such RBHs, the WEC is still preserved. Apart from
some matter fields, the WEC is satisfied by general matter fields.
According to Ref.~\cite{bala}, our solution should satisfy the WEC.
The WEC expresses that the local energy density of matter cannot be
negative for all local observers and dominates over the pressure
\cite{poi}. In other words, the energy density satisfies
$T_{\alpha\beta} u^\alpha u^\beta\geq0$, where $u^\alpha$ is a
timelike vector. This is considered to require an anisotropic fluid
in order to find a RBH. Thus, the energy-momentum tensor is given by
\begin{equation}
\label{mat:20}T^\alpha_{~\beta}=diag\left[
-\rho(r),p_r(r),p_\perp(r),p_\perp(r)\right],
\end{equation}
where $p_r=-\rho$ is the radial pressure. So, the energy-momentum
tensor confirms the symmetry $T^0_{~0} = T^1_{~1}$ as expected. The
tangential pressure $p_\perp$ is given by
\begin{equation}
\label{mat:21}p_\perp=-\rho-\frac{r}{2}\partial_1\rho,
\end{equation}
Instead of a point particle, a source turns into a magnetic droplet
of anisotropic fluid of density $\rho$. On physical grounds, a
non-zero radial pressure is equivalent to preventing droplet to
collapse into a matter point. This is an inherent effect on matter
owing to the spacetime noncommutativity and it is needed to balance
the inward gravitational pull.

The WEC can be expressed as the following inequalities
\begin{equation}\label{mat:22}  \Bigg\{
\begin{array}{ll}
\rho\geq0,\\
\rho+p_i\geq0;\quad i=1,2,3.\\
\end{array}
\end{equation}
The inequalities above in terms of the mass function can be written
as
\begin{equation}\label{mat:23}  \Bigg\{
\begin{array}{ll}
\frac{1}{r^2}\frac{dm(r)}{dr}\geq0,\\
\frac{2}{r}\frac{dm(r)}{dr}\geq\frac{d^2m(r)}{dr^2},\\
\end{array}
\end{equation}
where
\begin{equation}
\label{mat:24}\Bigg\{
\begin{array}{ll}
\frac{dm(r)}{dr}=m_g
\left(\frac{r^2\textmd{e}^{-\frac{r^2}{4\theta}}}{2\sqrt{\pi\theta^3}}+\frac{3g^3}{r^4}\frac{m}{M}\right),\\
\frac{d^2m(r)}{dr^2}=\frac{1}{r^3+g^3}\left[3g^3
\left(\frac{1}{r}\frac{dm(r)}{dr}-\frac{m}{r^2}-\frac{3mr}{r^3+g^3}\right)+\frac{Mr^4\textmd{e}^{-\frac{r^2}{4\theta}}}{2\sqrt{\pi\theta^3}}
\left(-\frac{3r^3}{r^3+g^3}-\frac{r^2}{2\theta}+5\right)\right].\\
\end{array}
\end{equation}
From above relations, it is easy to check that noncommutative
Hayward BHs satisfy the WEC everywhere. However, there are some RBH
solutions which do not satisfy the WEC
\cite{ayo2,ayo3,ayo5,bron,buri}. Here we should emphasize that the
authors in Ref.~\cite{bala} have shown that the construction of
charged RBH metrics is established upon three requirements: the
satisfaction of the WEC, the confirmation of the symmetry $T^0_{~0}
= T^1_{~1}$ and the asymptotic behavior of the solution as the R-N
BH metric. It is evident that our results agree with the mentioned
requirements found in \cite{bala}.

It is noteworthy that from the homogeneous equation $R_{\mu\nu} =
0$, any vacuum solution must be found only in the absence of matter.
However, the above homogeneous equation implicitly presumes that a
source term is concentrated on a point at the origin. It seems a
kind of physically inconsistent situation that curvature is produced
by a zero energy momentum tensor. In any pointless geometry, the
implementation of the homogeneous equation, which is dependent upon
the notion that matter is concentrated in a single point, is in
vain. This is indeed a proper motivation regarding the utility of
the noncommutative version of spacetimes. In the noncommutative
background, the energy momentum tensor is scattered throughout a
region of linear size determined by the parameter $\theta$.
Therefore, the virtue of the adopted approach lies in the fact that
contrary to other approaches, at very short distances, the
regularity of the spacetime is not artificial, but it immediately
emerges from the the fluid type picture of the quantum
noncommutative fluctuations, while at large distances, NCG turns
into a smooth classical manifold.

\section{\label{sec:4}Lagrangian uncertainty relation}
The HUP is widely considered to be an essential conceptual tool for
comprehending differences between classical and quantum theory. On
the assumption that the HUP is not always sufficient to confirm the
essence of what is non-classical about quantum mechanics, it is
sensible to speak of a generalization of it. In other words, a
complete formulation of the HUP may provide the quantum essence of
quantum theory. For example, the appearance of a minimal observable
length is a phenomenological feature of any approach to QG which
leads to the GUP in the following form \cite{taw}:
\begin{equation}
\label{mat:25}\Delta x\Delta p\geq\frac{1}{2}\left(1+\alpha(\Delta
p)^2+\ldots\right),
\end{equation}
where $\alpha$ is a constant of order unity (usually assumed
positive) and is dependent on the details of the QG theory. The GUP
relation signifies a finite minimal uncertainty $\Delta
x_0=\sqrt{\alpha}$. Hence, $\Delta x_0>0$ may be a manifestation of
the fuzzyness of space, or may be considered as a consequence of the
smeared structure of the fundamental particles. The above GUP also
implies the corresponding corrections to the commutation relation
between the momentum operator $\hat{p}$ and the position operator
$\hat{x}$ in the pertinent Heisenberg algebra as follows
\begin{equation}
\label{mat:26}[\hat{x},\hat{p}]=i(\hat{I}+\alpha \hat{p}^2+\ldots),
\end{equation}
where $\hat{I}$ shows the unit operator. These expressions show that
the structure of the spacetime is embellished with an effective
minimal length beyond which any measurements to observe a particle
location with a precision more than an intrinsic length scale is
impossible. The applications and consequences of such a minimal
length in a wide range of physical systems have been considered in
the literature \cite{GUP-Ali}.

In the light of the motivation mentioned above, as a starting point,
one can propose a similar idea with Ohanian and Ruffini \cite{oha}
but conversely to derive a kind of GUP. They utilized the HUP to
propose an alternative heuristic derivation of the Hawking
radiation. We may calculate a modified Hawking temperature and use
the heuristic viewpoint similar to Ref.~\cite{oha} to find a
modified uncertainty principle for radiated photons. On the basis of
thermodynamic consistency, BHs emit a thermal black body spectrum at
the Hawking temperature,
\begin{equation}
\label{mat:27}T_H=\frac{1}{4\pi}\frac{\partial_1\textsf{g}_{00}}{\sqrt{-\textsf{g}_{00}\textsf{g}_{11}}}\bigg|_{r=r_h}=
\frac{1}{4\pi}\frac{dN(r)}{dr}\bigg|_{r=r_h}.
\end{equation}
Using Eqs.~(\ref{mat:7}), (\ref{mat:10}) and the relation
$m(r_h)=\frac{r_h}{2}$ at the event horizon, one can easily find the
Hawking temperature of noncommutative Hayward BHs as follows
\begin{equation}
\label{mat:28}T_H=\frac{1}{4\pi r_h}-\frac{r_h}{8\pi}L(r_h).
\end{equation}
This temperature leads to a modified uncertainty relation. In this
setup, one can estimate the characteristic energy of the emitted
photons from the standard uncertainty principle. Near the BH
surface, there is an inherent uncertainty in the position of any
particle of about the horizon radius ($\Delta x\sim r_h$), due to
the behavior of its field lines \cite{adl2}, and also on dimensional
grounds. This leads to the momentum uncertainty in terms of the
distance uncertainty and Lagrangian density,
\begin{equation}
\label{mat:29}\Delta p\sim\frac{1}{2 r_h}-\frac{r_h}{4}L(r_h),
\end{equation}
and to the energy uncertainty $\Delta E \sim4\pi T_H$, which is
identified as the characteristic energy of the emitted photon or as
a characteristic temperature that agrees with the Hawking
temperature up to a factor of $4\pi$ as a calibration factor. From
the uncertainty principle, the following modified uncertainty
relation including a Lagrangian density function is given by
\begin{equation}
\label{mat:30}\Delta x\Delta
p\gtrsim\frac{1}{2}\left(1-\frac{(\Delta x)^2}{2}L(\Delta x)\right).
\end{equation}
The second term on the right-hand side of Eq.~(\ref{mat:30}) is
dependent on the parameters of Hayward and noncommutativity. The
term of corrections represents the nonlinear magnetic charge and
noncommutativity effects of probing photon rather than gravitational
effects in comparison with the GUP. The deviation from the standard
picture in the solution occurs only in a limited region around the
origin. In the limit $\Delta x\gg1$, the first term on the
right-hand side of the modified uncertainty relation is dominant and
the HUP is recovered, while in the regime $\Delta x\ll1$ the second
term is dominant and plays an essential role when the momentum and
distance scales are in the core around the origin, gravity is
actually described by a nonlinear electrodynamic field and the NCG
rather than by GR.

The characteristic Lagrangian density of emitted photons is
estimated from the HUP. Since $\Delta x$ is associated with the
horizon radius, we have $m(\Delta x)\approx\frac{\Delta x}{2}$.
Therefore, from Eq.~(\ref{mat:17}) we solve for the Lagrangian
uncertainty in terms of the distance uncertainty
\begin{equation}
\label{mat:31}\Delta L\approx\frac{2M}{(\Delta x)^3+g^3}
\left(\frac{(\Delta x)^3\textmd{e}^{-\frac{(\Delta
x)^2}{4\theta}}}{\sqrt{\pi\theta^3}}+\frac{3g^3}{M(\Delta
x)^2}\right).
\end{equation}
The Lagrangian uncertainty term emerges from the combined influence
of the nonlinear magnetic source and the noncommutativity, such that
for $L(\Delta x) \rightarrow 0$ the vacuum solution is recovered
that is as $\Delta x\Delta p\geq\frac{1}{2}$. Hence, when
$r\rightarrow\infty$ the noncommutative Hayward metric function
behaves as the Schwarzschild BH, which implies the HUP.

\begin{figure}[htp]
\begin{center}
\includegraphics[width=105 mm]{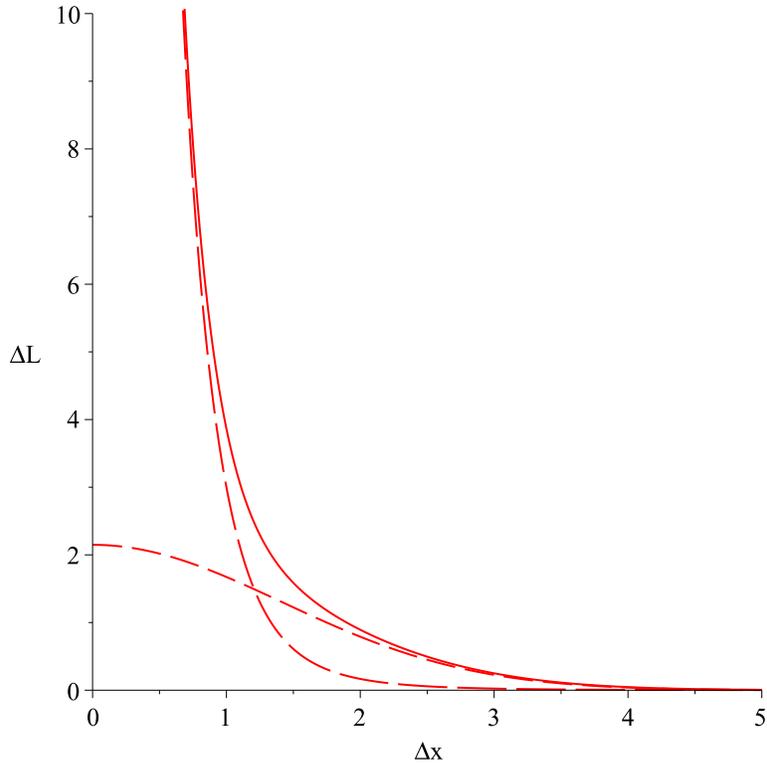}
\end{center}
\caption{\scriptsize {$\Delta L$ in terms of $\Delta x$ for specific
values of $g$ and $\theta$. The solid line corresponds to the
noncommutative Hayward BH for $g=1.00$ and $\theta=1.00$. The dash
line refers to the noncommutative Schwarzschild BH for $g=0$ and
$\theta=1.00$. The long dash line represents the Hayward BH for
$g=1.00$ and $\theta=0$. We have set $M=M(\textmd{min})$ for each
cases. }}
 \label{fig:2}
\end{figure}
For further specifications of this issue, the numerical results of
the Lagrangian uncertainty versus the distance uncertainty for
specific values of $g$ and $\theta$ are presented in
Fig.~\ref{fig:2}. As the figure shows, for the cases of Hayward and
noncommutative Hayward BHs we have roughly similar results at small
length scales ($\Delta x\sim1$), while the results of noncommutative
Schwarzschild and noncommutative Hayward BHs are substantially
similar at large length scales ($\Delta x\gg1$). However, the
results of three cases of BHs are equivalent at very large distances
($\Delta x\rightarrow\infty$), which leads to $\Delta
L\rightarrow0$. We can see that the behavior of the noncommutative
Schwarzschild case is rather different at very short distances. In
the limit $\Delta x\rightarrow0$, the Lagrangian uncertainty reaches
a maximum value for the noncommutative Schwarzschild BH, while it is
infinite for two other cases. Nevertheless, as previously stated, we
should consider the results just under the circumstance that $r \geq
r_0$. In other words, there is a minimum uncertainty in position
that is equivalent to the minimal nonzero radius, i.e. $\Delta
x(\textmd{min})\sim r_0$, such that it is impossible to set up a
measurement to find a more precise particle position than $\Delta
x(\textmd{min})$. As a result, the appearance of a lower finite cut
off at the short scale gravity compels a bound on any measurements
to determine a particle position in a noncommutative gravity theory.
Therefore, one should set the requirement that $ L(\Delta
x)\lesssim\Delta L(\textmd{max})$ because for the pattern of the
metric for $ \Delta L>\Delta L(\textmd{max})$ no sensible
temperature can be determined. This upper bound on the Lagrangian
uncertainty is a necessary condition for preventing the unusual
thermodynamical behavior at very short distances. In
Table~\ref{tab:1}, the numerical results of the minimum distance
uncertainty, minimum mass and maximum Lagrangian uncertainty for
specific values of $g$ and $\theta$ are presented. As the table
shows, the nonlinear electrodynamic field and/or the fictitious
charge lead to the existence of a remnant mass ($M_0$) in which the
BH can shrink to. The results are confirmed by the numerical
solutions of Ref.~\cite{meh2} for the three models of nonsingular
BHs. This means that the existence of the fictitious-magnetic
charge, fictitious charge and magnetic charge are responsible for
the nonsingular solutions in the noncommutative Hayward BH,
noncommutative Schwarzschild BH and Hayward BH, respectively. As a
final note, if we had chosen the Bardeen solution, as another
popular example of RBHs, despite that only the mass function would
have altered, the general properties
would have directed to wholly equivalent results to those above \cite{meh1}.\\

\begin{table}
\caption{\scriptsize This table shows the remnant radius, remnant
mass and maximum Lagrangian uncertainty for specific values of $g$
and $\theta$ that correspond to various kinds of BHs and hence are
in agreement with Fig.~\ref{fig:2}.}
\begin{center}
\begin{tabular}{|c|c|c|}
\hline          Noncommutative Hayward BH         & Noncommutative Schwarzschild BH            & Hayward BH   \\
\hline $\Delta x(\textmd{min})\sim r_0\approx3.13$& $\Delta x(\textmd{min})\sim r_0\approx3.02$& $\Delta x(\textmd{min})\sim r_0\approx1.26$   \\
\hline $ M(\textmd{min})\sim M_0\approx1.97$      & $ M(\textmd{min})\sim M_0\approx1.90$      & $M(\textmd{min})\sim M_0\approx0.94$   \\
\hline $ \Delta L(\textmd{max})\approx0.20$       & $\Delta L(\textmd{max})\approx0.22$        & $\Delta L(\textmd{max})\approx1.26$   \\
\hline
\end{tabular}
\end{center}
\label{tab:1}
\end{table}

\section{\label{sec:5}Conclusions}
In summary, we have investigated a general ansatz of the BH solution
having the static spherical symmetric configuration with a
fictitious charge and the nonlinear magnetic charge in the context
of Einstein-NED theory. We have studied the Hawking temperature of
noncommutative Hayward BHs. In this way, a modified uncertainty
relation is achieved in a heuristic way by the utilization of the
Hawking temperature and general properties of BHs according to
Ref.~\cite{oha}. The term of corrections is due to the Hayward's
parameter, stemming from a nonlinear electrodynamic field, plus
noncommutativity effects. As a result, the modified uncertainty
relation including a Lagrangian density function is emerged from
nonlinear electromagnetism as a physical source and an inherent
property of the manifold instead of gravitational effects compared
to the GUP. Ultimately, it is found that there is an upper bound on
the Lagrangian uncertainty of BHs at the short scale gravity which
comes from the deSitter core around the origin.

The authors in Ref.~\cite{ali1} have recently shown that the
behavior of Hawking radiation close to the Planck phase is
significantly modified in the framework of gravity's rainbow. This
modification gives corrections to the thermodynamic description of
BHs and predicts the existence of BH remnant. As a potential
problem, it will be interesting to investigate the behavior of
Hawking radiation for the noncommutative inspired Hayward BH through
rainbow functions.

In addition, it is important to point out the very recent study on
the noncommutative inspired electrically and magnetically charged BH
in Euler-Heisenberg NED model \cite{Maceda:2019}. In this model, the
WEC was examined and satisfied in opposition to the commutative
case. In their setup, the formation of shadows for the
noncommutative inspired metrics was addressed and also it was shown
that the modification due to the noncommutativity might be
susceptible to observation using new generation interferometers. It
would also be interesting to extend the present analysis to the
Euler-Heisenberg NED model.

Finally, it is interesting to note that the Cauchy horizons of the
R-N BHs \cite{RN}, Schwarzschild white holes \cite{SWH} and
Schwarzschild wormholes \cite{SW} are unstable. In general, an
observer crossing the Cauchy horizon might encounter an arbitrarily
large blue shift of any incoming radiation; so that any small
perturbation might disrupt the horizon and result in a curvature
singularity. In what concerns the possible blue shift instability at
the inner Cauchy horizon of a noncommutative inspired BH, one should
emphasize that the curvature singularity is relevant only for a
classical differential manifold. The propagation of any field in a
noncommutative spacetime is under the influence of a natural
ultraviolet cutoff in which an infinite amount of energy near the
inner Cauchy horizon cannot be observed. On the other hand, the
violation of the strong energy condition could occur outside the
inner horizon wherein gravity is actually described by NCG and not
by GR. Thereby, the situation for noncommutative inspired BHs is
quite different and we leave it as an open problem that goes beyond
the scope of the present paper. We hope to further investigate this
issue in future work.

\end{document}